\documentstyle[12pt]{article}
\begin{document}
\begin{center}{\Large {\bf Form invariant Sommerfeld 
Electrical Conductivity
in generalised d - dimensions}}\end{center}

\vskip 1cm

\begin{center}{\it Muktish Acharyya}\\
{\it Department of Physics,}\\
{\it Presidency University, 86/1 College Street,}\\
{\it Calcutta-700073, India.}\\
{\it E-mail:muktish.acharyya@gmail.com}\end{center}

\vskip 1.5cm

\noindent {\bf Abstract:} The Sommerfeld electrical conductivity is
calculated in {\it d} dimensions following Boltzmann kinetic approach. At
$T=0$, the mathematical form of the electrical conductivity is found to
remain invariant in any generalised spatial ($d$) dimensions. 

\vskip 1.5cm


\newpage

The behaviour of conduction electrons plays a major role to govern
the properties of metallic solids \cite{mott,wilson,dekker,ashcroft}.
The electronic specific heat, the spin paramagnetism, the electrical
conductivity etc., are some examples of these properties. 
Actually, in the solids,
the density of electronic states plays the key role. Theoretically,
the electronic behaviours are studied by using the density of states
(for free electrons) in two and three dimensions. The distributions
of electrons in these states are made by using the 
Fermi-Dirac function \cite{huang,schroeder,pathria}.
All these calculations are done (and available in standard literature)
mainly in two and three dimensions, since the experimental results are
available in these dimensions. However, one may extend these studies in
generalised dimensions also just for pure pedagogical reason. For example,
the behaviours of free electrons in generalised dimensions \cite{ma1}.
As a result, in infinite dimensions, 
{\it interestingly}, it was found that the fermi level is
populated by all electrons \cite{ma1} 
without violating Pauli exclusion
principle !
Another example is the
temperature variation of Pauli spin susceptibility. The general expression
for this susceptibility was derived in {\it d} dimensions \cite{ma2}.
From this expression, it was found that the Pauli susceptibility becomes
temperature independent {\it only} in two dimensions \cite{ma2}. All
these studies, although pedagogical, give very interesting results. 

One may try to generalise the behaviour of electrical conductivity
\cite{som1,som2} in generalised {\it d} dimensions. As far as the
knowledge of this author is concerned, this type of calculation
is not available in standard literatures of condensed matter physics.
If we see the expression of electrical conductivity (derived by
Sommerfeld\cite{som3}) in three dimensions, the form is given 
as $\sigma =
 {{ne^2\tau} \over {m}}$. Where $\tau$ is the relaxation time of the
system. It is not yet known, whether this form will
remain invariant in any other dimensions. 

In the present article, I have rederived the expression of electrical
conductivity at $T=0$, 
in generalised {\it d} dimensions following Boltzmann kinetic
approach. 

In the relaxation approximation, the Boltzmann equation takes
the form\cite{dekker}
\begin{equation}
({{\partial f} \over {\partial t}})_{collision} = - {{f - f_0} \over \tau}
\end{equation}

again, in the presence of electric and magnetic field,
\begin{equation}
({{\partial f} \over {\partial t}})_{collision} = 
-e \left( \vec E + {{{\vec v} \times {\vec H}} \over c} \right) \cdot 
{\nabla}_p f + {\vec v} \cdot {\nabla}_r f.
\end{equation}

For free electrons, the nonequilibrium 
distribution function $f$ is function of the
electronic momenta only. 
In {\it d} dimensions, it may be represented as $f(p_1,p_2,p_3,...p_d)$.
It does not depend on the electronic position 
coordinates. So, ${\nabla}_r f = 0$. In the absence of any magnetic
field $\vec H = 0$. If the electric field is applied 
only along a particular
direction (say direction 1), then
\begin{equation}
({{\partial f} \over {\partial t}})_{collision} = 
-e {E_1} {{\partial f} \over {\partial {p_1}}}
\end{equation}
So, (from (1) and (3)), one may write,
\begin{equation}
{{f-f_0} \over {\tau}} = 
eE_1 {{\partial f} \over {\partial {p_1}}}
\simeq
eE_1 {{\partial {f_0}} \over {\partial {p_1}}}
\end{equation}
Here, it is assumed (relaxation or linear approximation) 
that the electronic nonequilibrium 
distribution function $f$ does not change
too much from its equilibrium distribution function $f_0$. In generalised
$d$ dimensions, the total energy (kinetic) of a single electron may be
expressed as
\begin{equation}
\epsilon = (p_1^2 + p_2^2 + p_3^2 + .....+ p_d^2)/2m
\end{equation}
It may be noted here, that this is an equation of $d$ dimensional 
hypersphere of radius $R=(2m\epsilon)^{1/2}$.

The term ${{\partial {f_0}} \over {\partial {p_1}}}$ is equal to 
$({{\partial {f_0}} \over {\partial {\epsilon}}})$  
$({{\partial {\epsilon}} \over {\partial {p_1}}})$.  
And
$({{\partial {\epsilon}} \over {\partial {p_1}}}) = p_1/m = v_1$.  
So,
\begin{equation}
{{f-f_0} \over {\tau}} = 
eE_1v_1 {{\partial {f_0}} \over {\partial \epsilon}}
\end{equation}
In $d$ dimensions the current density may be written as
\begin{equation}
I_1 = -{{2e} \over {h^d}} \int \int...\int v_1 (f-f_0)dp_1dp_2....dp_d
\end{equation}

Using eqn (6) one may write,
\begin{equation}
I_1 = -{{2e^2} \over {h^d}} E_1\int \int...\int v_1^2 
\tau(\epsilon)({{\partial {f_0}} \over 
{\partial \epsilon}})dp_1dp_2....dp_d.
\end{equation}
Assuming the relaxation time $\tau$ is function of electronic energy
$\epsilon$ only. 
$d^dp=dp_1dp_2....dp_d$ is the elementary volume in $d$
dimensional hypersphere (Ref. \cite{pathria}, Appendix [c]). This is
calculated and is given as
\begin{equation}
d^dp = 
({d \over 2}){{\pi^{d/2}} \over {\Gamma(d/2+1)}} (2m)^{d/2}
{\epsilon}^{{d-2} \over 2} d\epsilon.
\end{equation}
Considering the system to be isotropic i.e., $v_1^2 = v^2/d = {{p^2}
\over {m^2d}}= {{2m\epsilon} \over {m^2d}} = {{2\epsilon} \over {md}}$.
Now, the equation (8) takes the form
\begin{eqnarray}
I_1 
&=& -{{2e^2} \over {h^d}} E_1
({d \over 2}){{\pi^{d/2}} \over {\Gamma(d/2+1) md}} (2m)^{d/2}
2\int \epsilon 
\tau(\epsilon)({{\partial {f_0}} \over {\partial \epsilon}})
\epsilon^{{d-2} \over 2} d\epsilon \nonumber\\
&=& {{2e^2} \over {h^d}} E_1
{{\pi^{d/2}} \over {\Gamma(d/2+1) m}} (2m)^{d/2}
{\epsilon_F}^{d/2}\tau(\epsilon_F) 
\end{eqnarray}
Here, in the last step, the relation ${{\partial {f_0}}
\over {\partial {\epsilon}}} = -\delta(\epsilon - \epsilon_f)$ is
used.

The number of electronic states in unit spatial volume and momentum
between $\vec p$ and $\vec p + \vec {dp}$ is $2d^dp/h^d$. The factor
2 comes due to the fact that 
each momentum states has two fold spin degeneracy
for the electrons.
\begin{eqnarray}
{{2} \over {h^d}}d^dp 
&=& g(\epsilon) d\epsilon\nonumber\\
&=& 
{{2} \over {h^d}} 
({d \over 2}){{\pi^{d/2}} \over {\Gamma(d/2+1)}} (2m)^{d/2}
\epsilon^{{d-2} \over 2}d\epsilon
\end{eqnarray}

where $g(\epsilon)d\epsilon$ is the number of electronic states between
energy $\epsilon$ and $\epsilon+d\epsilon$. The number of electrons per
unit volume at $T=0$ can be calculated as
\begin{eqnarray}
n &=&
 \int_0^{\infty} g(\epsilon)
 f_0 (\epsilon) d\epsilon\nonumber\\
&=& 
 \int_0^{\epsilon_F} g(\epsilon)
 d\epsilon\nonumber\nonumber\\
&=&
{{2} \over {h^d}} 
{{\pi^{d/2}} \over {\Gamma(d/2+1)}} (2m)^{d/2}
(\epsilon_F)^{d/2}.
\end{eqnarray}

Where, in the last step the following form fermi function ($T=0$) is used

\begin{eqnarray}
f_0(\epsilon) &=& 1 ~~~~{\rm for ~~~\epsilon \leq \epsilon_F} \nonumber\\
&=& 0 ~~~~{\rm for ~~~\epsilon > \epsilon_F}. \nonumber
\end{eqnarray}

Using eqn (12), the eqn (10) becomes

\begin{equation}
I_1 = E_1 {{ne^2\tau(\epsilon_F)} \over m}
\end{equation}

The electrical conductivity becomes

\begin{equation}
\sigma = I_1/E_1 = {{ne^2\tau(\epsilon_F)} \over m} 
\end{equation}

The same result was obtained by 
Sommerfeld in three dimensions\cite{som3}.
 The same form
of electrical conductivity is obtained here in the case of generalised
$d$ dimensions. {\it So, the form of Sommerfeld electrical conductivity
remains invariant in generalised d dimensions}.

In the concluding remarks, 
I would like to mention that unless one calculate
these electronic physical quantities in generalised {\it d} dimensions,
some subtle theoretical important informations would be missing. 
For example,
the temperature dependence of Pauli spin susceptibility  
was already 
known\cite{dekker} (in $d=3$), 
but it's temperature independence in two dimensions {\it only}, is
the new result\cite{ma2}. 
This was possible to derive only due to the generalisation
in {\it d} dimensions. 
In this article, the {\it invariant form} of the Sommerfeld
electrical conductivity, was possible to derive only due to the 
generalisation in {\it d} dimensions. 

Since, these results are obtained in {\it d} dimensions, for ${
d > 3}$, one cannot compare the theoretical results with that obtained
experimentally. In fact, it is not really known, being embedded in
three dimensional realistic space
how one can simulate
the dimensionality of the electronic system, above three. 
So, right now, these results cannot be compared with experimental
results.
One thing
may be mentioned here that, since the form of electrical conductivity,
remains invariant in any generalised dimensionality, 
the thin wires ($d=1$),
thin films ($d=2$) and bulk materials ($d=3$) will show similar
{\it forms} of electrical conductivities. Only the measures of relaxation
time will be different. It may also be noted that, these calculations are valid only at T=0.
The derivative of Fermi function used here, is Dirac delta type. That is another reason
that this cannot be compared with experimental result.

Some diluted metal composites has fractional dimensionality near
the percolation threshold for electrical conduction. The results
presented in this paper will not be able to simulate those metal
composites. The reason is that
the way of getting the fractal dimension is different
from that for integer dimension. Another point may be mentioned
in this context that, the fractal
dimension of metal composites, is always less than three. 

\vskip 0.5cm

\noindent {\bf Acknowledgements:} Author would like to thank N. Banerjee
and P. Rudra for important discussions.

\end{document}